\documentclass[12pt]{article}
\usepackage{amsmath,amssymb,graphicx}
\usepackage{color}
\usepackage{epsfig}
\usepackage{amssymb}
\usepackage{float}
\usepackage{extarrows}
\newcommand{\be}{\begin{equation}}
\newcommand{\bea}{\begin{eqnarray}}
\newcommand{\eea}{\end{eqnarray}}
\newcommand{\ba}{\begin{array}}
\newcommand{\ea}{\end{array}}
\newcommand{\ee}{\end{equation}}

\newcommand{\lan}{\langle}
\newcommand{\ran}{\rangle}
\newcommand{\lam}{\lambda}
\begin{document}

\title{\textbf{Cachazo-Svr\v{c}ek-Witten Rules for Tree-Level Gluonic Amplitudes Revisited}}
\author{Wen-Jie Zhang$^1$,~~
Jun-Bao Wu$^{1, 2}$\footnote{Corresponding author. E-mail: junbao.wu@tju.edu.cn},~~
Chuan-Jie Zhu$^3$~}

\maketitle

\begin{center}
{\it
$^{1}$Center for Joint Quantum Studies and Department of Physics, School of Science,\\
Tianjin University, 135 Yaguan Road, Tianjin 300350, P.~R.~China\\
$^{2}$Peng Huanwu Center for Fundamental Theory, Hefei, Anhui 230026, P. R. China \\
$^{3}$College of Mathematics and Physics Science,\\ Hunan University of Arts and Science, Changde
415000, P. R. China
}
\vspace{10mm}
\end{center}


\begin{abstract}
We provide a new proof of Cachazo-Svr\v{c}ek-Witten rules for tree-level gluonic amplitudes. As a key step, we explicitly demonstrate the cancellation of spurious poles originating from the maximally helicity violating vertices in these rules. To achieve this, we introduce specially-defined two-off-shell-line sub-amplitudes and examine their residues at spurious poles.\\

PACS codes: 11.15.-q, 11.15.Bt, 12.38.Bx.\\

Keywords: General properties of perturbation theory, Perturbative calculations, Dualities in gauge field theories.
  \end{abstract}
\baselineskip 18pt

\section{Introduction}
Great achievements \cite{Henn:2014yza, Elvang:2015rqa, Arkani-Hamed:2012zlh} on scattering amplitudes in gauge theories in the past two decades were largely inspired by the seminal paper of Witten on the relationship between amplitudes and twistor string theory \cite{Witten:2003nn}. Tree-level maximally helicity violating (MHV) amplitudes were reproduced \cite{Witten:2003nn} from integration over moduli space of genus-zero degree-one curves\footnote{These curves are $D$-instatons in $B$ model with target space ${\mathbb P}^{3|4}$.} in super twistor space ${\mathbb P}^{3|4}$. The tree non-MHV amplitudes are corresponding to genus-zero curves with degrees greater than one. More precisely, $\text{N}^k\text{MHV}$ amplitudes (amplitudes with $k+2$ negative-helicity gluons) are related to curves with degree $d=k+1$. Later it was discovered that for non-MHV tree amplitudes, one can either integrate over the moduli space of  connected curves with degree $d$ or over the moduli space of $d$  degree-one curves (completely disconnected instantons). The former choice led to the compact Roiban-Spradlin-Volovich formula \cite{hep-th/0402016, hep-th/0402121, hep-th/0403190}, while the latter resulted in the Cachazo-Svr\v{c}ek-Witten (CSW) rules for tree amplitudes \cite{Cachazo:2004kj}. The equivalence of these two formulations was demonstrated in \cite{hep-th/0404085}. Intermediate prescriptions using integrate over moduli space of $m \, (1<m<d)$ curves with total degree $d$ were also proposed in \cite{hep-th/0404085}. Twistor space recursion relations for these intermediate prescriptions were given in \cite{hep-th/0406133}.

CSW introduced MHV vertices as an off-shell continuation of MHV amplitudes. Together with scalar propagators, these MHV vertices form MHV diagrams. Using the MHV diagrams to calculate tree amplitudes is more efficient than using the Feynman diagrams. For reviews including CSW rules, see \cite{Cachazo:2005ga, Brandhuber:2011ke, Feng:2011np}.  The analysis of various physical and spurious singularities in \cite{Cachazo:2004kj}, combined with the dimension analysis (or on-shell recursion relations) in \cite{Britto:2005fq}, provided the first proof of the CSW rules.  This proof includes three key steps. The first is to demonstrate that an amplitude derived from CSW rules, $A_{\text{CSW}}$, has correct collinear and multi-gluon singularities. The second step is to demonstrate that all spurious poles related to the off-shell continuation are cancelled among themselves. This result implies that $A_{\text{CSW}}$ is Lorentz covariant \cite{Cachazo:2004kj}.  From the first two steps, we get that the difference between $A_{CSW}$ and $A_{\text{Feynman}}$, which is the amplitude from Feynman rules, can only be at most polynomials of $\lambda$'s and $\tilde\lambda$'s.  The last step is shown this polynomial should be zero by using  the fact that any $n$-gluon tree amplitude, either from CSW rules or Feynman rules,  has mass dimension $4-n$ and direct computations in the case with $n=4$. This step can be replaced by showing that both $A_{\text{CSW}}$ and the amplitude from Feynman rules, $A_{\text{Feynman}}$, satisfy the same Britto-Cachazo-Feng-Witten (BCFW) recursion relation \cite{Britto:2005fq}.

The second step of the proof was displayed in detail for tree next-to-MHV (NMHV) amplitudes involving MHV diagrams with a single propagator \cite{Cachazo:2004kj}.  This discussion can be generalized for N$^\text{k}$MHV amplitudes, but the proof seems quite complicated when $k$ is large.  NMHV amplitudes are special, not only because each MHV diagram for them always has exactly one propagator, but also the spurious poles in such amplitudes are always degenerate according to classification we will provide later in this paper. In this paper, we would like to provide another treatment for this step, based on a generalization of the treatment in \cite{Zhu:2004kr, Wu:2004fba}  for googly  ($\overline{\text{MHV}}$) amplitudes. At first glance, the treatment there appears to be heavily reliant on the unique properties of MHV diagrams for googly amplitudes: there is only one MHV vertex with four lines, and all other vertices have exactly three lines. As we will demonstrate in the main body of this paper, the treatment here can still be non-trivially generalized to the most general cases to demonstrate that spurious poles are cancelled among themselves.
Risager provided another proof \cite{hep-th/0508206} of CSW rules based on multi-gluon BCFW-like shifts which resulted in new on-shell recursion relation\footnote{A proof of CSW rules for all tree amplitudes in ${\cal N}=4$ super Yang-Mills theory using all-line shift was provided in \cite{0811.3624}. Derivation of CSW rules from Lagrangian was studied in \cite{hep-th/0510111, hep-th/0511264, hep-th/0605121, hep-th/0611164, hep-th/0703286} and a twistor-action formulation was provided in \cite{hep-th/0702035}.}. Our approach here is complementary to others. We hope that our proof will provide new insight into CSW-like rules for amplitudes in various other theories.

In the following section, we will provide a brief introduction to the CSW rules. Section \ref{section:proof} is the main part of this paper and presents our proof of the CSW rules. First, we introduce specially-defined two-off-shell-line sub-amplitudes and demonstrate that their residues at related spurious poles are proportional to one-off-shell-line sub-amplitudes. Based on this result, in subsection \ref{subsection:cancellation}, we demonstrate that all spurious poles cancel among themselves by dividing them into groups. In subsection \ref{subsection:completion} we complete our proof based on some known facts. Section \ref{section:conclusion} is devoted to a conclusion and some discussions on possible further applications of our approach. Some technical details are put in several appendices.

\section{Brief introduction to CSW rules based on MHV diagrams}

Our conventions follow closely \cite{Witten:2003nn}, especially the signature is chosen to be $(+---)$. By using the crossing symmetry, we make all momenta of external gluons outgoing. For the momentum $p_\mu$ carrying by an external gluon, the following decomposition into  bispinors will be used,
\be\label{bispinor} p_{\alpha \dot{\alpha}}\equiv p_\mu\sigma^\mu_{\alpha\dot{\alpha}}=\lambda_\alpha\tilde{\lambda}_{\dot{\alpha}}\,,\ee
where $\sigma^\mu=(\mathrm{1}_{2\times 2}, \sigma^i)$, with $\mathrm{1}_{2\times 2}$ two-dimensional identity matrix and $\sigma^i$'s three Pauli matrices.
The spinor index $\alpha$ ($\dot{\alpha}$) will be raised (or lowered) by anti-symmetric tensor $\epsilon^{\alpha\beta}$ ($\epsilon^{\dot{\alpha}\dot{\beta}}$) and $\epsilon_{\alpha\beta}$ ($\epsilon_{\dot{\alpha}\dot{\beta}}$),
\be\lambda^\alpha=\epsilon^{\alpha\beta}\lambda_\beta,\, \lambda_\alpha=\epsilon_{\alpha\beta}\lambda^\beta\,, \ee
with $\epsilon^{\alpha \beta}\epsilon_{\beta\gamma}=\delta^\alpha_\gamma$.\footnote{We explicitly choose $\epsilon_{12}=-\epsilon_{21}=-\epsilon^{12}=\epsilon^{21}=1$ for $\epsilon_{\alpha\beta}, \epsilon^{\alpha\beta}, \epsilon_{\dot{\alpha}\dot{\beta}}, \epsilon^{\dot{\alpha}\dot{\beta}}$.} We omit similar expressions involving $\tilde{\lambda}^{\dot{\alpha}}$ and $\tilde{\lambda}_{\dot{\alpha}}$. The following Lorentz invariant anti-symmetric inner products,
 \be\langle \lambda, \mu \rangle=\epsilon_{\alpha\beta}\lambda^\alpha\mu^\beta\,,\ee
 \be[\tilde{\lambda}, \tilde{\mu}]=\epsilon_{\dot{\alpha}\dot{\beta}}\tilde{\lambda}^{\dot{\alpha}}\tilde{\mu}^{\dot{\beta}}\,, \ee
 are quite useful.
We will also denote $\langle \lambda_r, \lambda_s\rangle$ ($[\tilde{\lambda}_r, \tilde{\lambda}_s]$) by $\langle r, s\rangle $ ($[r, s]$).

We will always focus on the partial amplitudes $A(1, 2, \cdots, n)$ with color factor stripped. These partial amplitudes are defined by the color decomposition of full amplitudes ${\cal M}(1, 2, \cdots, n)$ as
\be {\cal M}(1, 2, \cdots, n)=\sum_{\sigma\in S_n/{\mathbf Z}_n} \mathrm{Tr}(T^{a_{\sigma(1)}}T^{a_{\sigma(2)}}\cdots T^{a_{\sigma(n)}})A(\sigma(1), \sigma(2), \cdots, \sigma(n))\,, \ee
where   $a_{i}$'s are color indices of the $i$th gluon, and the summation is over the permutation group $S_n$ acting on the $n$ external gluons modulo the cyclic group ${\mathbf Z}_n$ acting as cyclic permutations. Without losing generality, we choose the gauge group to be $U(N)$. $T^{a_{i}}$'s are representation matrices (in the fundamental representation) of the generators of the gauge group, with normalization $\text{Tr}\left({T^{a}T^{b}} \right) = \delta^{ab}$.

From the  color decomposition, we know that we only need to compute the  partial amplitude\footnote{From now on, we will often use `amplitudes' to mean `partial amplitudes'.} $A(\sigma(1), \sigma(2), \cdots, \sigma(n))$. However, this is still a very complicated object because even for small $n$ like $n=5$, it contains a large number of possible Lorentz invariant combinations of momenta and polarization vectors (see, for example, Fig.~7 of \cite{Bern:1992ad}). Because we have introduced bi-spinor representation for external on-shell momenta $p_i=\lambda_i\tilde\lambda_i$, a partial amplitude is specified by $\lambda$, $\tilde{\lambda}$ and helicity $h$ of each gluon. We then denote a partial amplitude simply by $A_{n} \left(1^{h_{1}}, \ldots, n^{h_{n}} \right)$. To achieve this, polarization vectors are constructed as \cite{Xu:1986xb, Berends:1987cv}
\be
	\varepsilon_{\alpha\dot{\alpha}}^{(-)} = \frac{\lambda_{\alpha} \tilde{\mu}_{\dot{\alpha}}}{[ \tilde{\lambda}, \tilde{\mu}]}, \quad  \varepsilon_{\alpha\dot{\alpha}}^{(+)} = \frac{\mu_{\alpha} \tilde{\lambda}_{\dot{\alpha}}}{\langle \mu, \lambda\rangle},\label{eq:polarizations}
\ee
where $\mu$ and $\tilde{\mu}$ are reference spinors. The superscripts of polarization vectors $(+)$, $(-)$ denote positive helicity and negative helicity, respectively. $\mu$ ($\tilde{\mu}$) can vary depending on the external gluon because of gauge invariance, the final amplitude results are unaffected by the reference spinors used. The amplitude computation is greatly simplified by this spinor helicity trick.

We always refer to tree amplitudes with exactly two negative-helicity gluons when we talk about tree-level MHV amplitudes. The results for the $n$-gluon MHV amplitude is \cite{Parke:1986gb, Berends:1987me}
\be A_n^{\text{MHV}}(1^+, \cdots, r^-, \cdots, s^-, \cdots, n^+)=\frac{\langle r, s \rangle^4}{\prod_{i=1}^n \langle i, i+1 \rangle}. \ee

CSW \cite{Cachazo:2004kj} proposed a new method for computing tree-level amplitudes in Yang-Mills theory by using tree MHV diagrams with whole MHV amplitudes treated as vertices with suitable off-shell continuation (called MHV vertices). Some legs will be internal because the vertices will be inserted in MHV diagrams. The question about how to define $\lambda$ for such momentum arises.  CSW considered the following off-shell continuation for internal gluon with momentum $p$
\be
	\lambda_{\alpha} = p_{\alpha \dot{\alpha}} \tilde{\eta}^{\dot{\alpha}},
\ee
where $\tilde{\eta}^{\dot{\alpha}}$ is an arbitrary spinor that remains constant for all of the internal lines in all diagrams contributing to a given partial amplitude. For external gluons, the associated spinors are still obtained from the decomposition in \eqref{bispinor}. 

To construct MHV diagrams, we connect MHV vertices with propagators. We need to assign a factor $1/p^{2}$ for each propagator with internal momentum $p$. It is worth noting that the helicities at the two ends of each propagator are opposite. Each MHV diagram's contribution is provided by  the product of MHV vertices and scalar propagators. The tree-level MHV diagram must be planar, with the cyclic order of external line particles consistent with the cyclic order of external line particles in the to-be-computed partial amplitude. Partial amplitudes are obtained by summing contributions from all tree MHV diagrams with the correct cyclic order of external gluons.
One of the key features of the CSW rules is that the final result is independent of $\tilde{\eta}$. The crucial step of our proof here is to show this by generalizing some steps in the computations of googly amplitudes \cite{Zhu:2004kr, Wu:2004fba}. Our treatment is different from the one in \cite{Cachazo:2004kj}.

\section{Proof of the CSW rules\label{section:proof}}
\subsection{Two-off-shell-line sub-amplitudes and their spurious poles\label{subsection:colllinear}}

Spurious poles in MHV diagrams are caused by the off-shell continuation of MHV amplitudes.
The analysis of spurious poles in $A_{\text{CSW}}$ plays an important role in our proof. These poles appear as a result of the off-shell continuation in the CSW rules. From this subsection, we begin our study of the residues of these poles.

For our convenience, we introduce the following notation\footnote{\label{footnote1}When $a>b$, $\sum_{i=a}^b$ always means $\sum_{i=a}^n+\sum_{i=1}^b$.}
\be p_{a, b}\equiv\sum_{i=a}^b p_i\,, \ee
and as reviewed in the previous section, 
\be
p_{i\alpha\dot{\alpha}} = p_{ i\mu}\sigma^{\mu}_{\alpha\dot{\alpha}}=\lambda_{i\alpha}\,
\tilde\lambda_{i\dot\alpha},
\ee
for on-shell external momentum $p_{i}$. For generic external momenta $p_i$'s, $p_{a,b}$ is off-shell for $a\neq b$ and $b\neq a-2$. For this $p_{a,b}$, CSW \cite{Cachazo:2004kj} introduced a (holomorphic) $\lambda$:
\be
\lambda_{p_{a, b}}=\sum_{i=a}^b\lambda_i\phi_i, \quad  	\phi_{i} \equiv \tilde{\lambda}_{i\dot{\alpha}} \tilde{\eta}^{\dot{\alpha}}, \ee
where $\tilde\eta$ is a generic auxiliary (anti-holomorphic) spinor. According to the CSW rules \cite{Cachazo:2004kj}, this $\lambda_{p_{a,b}}$ will be used in the internal MHV vertex with an internal line of momentum $p_{a,b}$.

For generic external momenta $p_i$'s, we assume that any  two different $p_{a,b}$'s
corresponding $\lambda$'s are not  proportional to each other for generic $\tilde\eta$:
\be
\lambda_{p_{a,b}} \not\propto \lambda_{p_{c,d}} ,
\ee
except the special case: $p_{c,d}=p_{b+1,a-1}=-p_{a,b}$ 
Of course, $\lambda_{p_{a, b}}$ is not proportional to any $\lambda_i$'s.

Possible spurious  poles  appear when $\langle\lambda_{p_{n_1, n_2-1}}, \lambda_{p_{n_2, n_3-1}}\rangle$ (with $n_{i+1}-n_i>1, i=1, 2$)\footnote{The label $n_i$ should be understood as modulo $n$, as in footnote \ref{footnote1}.} or $\langle\lambda_{p_{n_1, n_2-1}}, \lambda_{n_2}\rangle$ vanishes and appears in the denominator of the expression for an MHV vertex.

We refer to the cases with two $\lambda$'s from off-shell momenta, like $\langle\lambda_{p_{n_1, n_2-1}}, \lambda_{p_{n_2, n_3-1}}\rangle$ with $n_2-n_1>1, n_3-n_2>1, n_1-n_3>1$ as non-degenerate  and  all other cases as degenerate.

For definite $n_1$, $n_2$ and $n_3$, let us define the momenta $P_1, P_2, P_3$,
\bea
P_1&\equiv&p_{n_1, n_2-1},\\
P_2&\equiv&p_{n_2, n_3-1},\\
P_3&\equiv&p_{n_3, n_1-1},
 \eea
and the spinors $v_1, v_2, v_3 $ to be
\bea
	 v_{1} &\equiv&\lambda_{P_1}= \lambda_{p_{n_{1}, n_{2}-1}} = \sum_{i=n_{1}}^{n_{2}-1} \lambda_{i}\phi_{i}\,, \\
	 v_{2} &\equiv&\lambda_{P_2}= \lambda_{p_{n_{2}, n_{3}-1}} = \sum_{i=n_{2}}^{n_{3}-1} \lambda_{i}\phi_{i}\,, \\
	v_{3} &\equiv&\lambda_{P_3}= \lambda_{p_{n_{3}, n_{1}-1}} = \sum_{i=n_{3}}^{n_{1}-1} \lambda_{i}\phi_{i}\,.
\eea
It is easy to see that momentum conservation leads to
\be\sum_{i=1}^3v_i=0\,,\ee
which leads to
\be\lan v_1, v_2 \ran=\lan v_2, v_3\ran=\lan v_3, v_1\ran\,. \ee

When $\langle v_1, v_2 \rangle=0$, there exist $\alpha_1, \alpha_2 \in \mathbf{C}$ and spinor $v_0$ such that
\be v_i=\alpha_i v_0, \, i=1, 2\,.\ee
Then $v_3=-(\alpha_1+\alpha_2)v_0$. We now define $\alpha_3$ to be $\alpha_3\equiv-(\alpha_1+\alpha_2)$.

For an arbitrary function of $v_1, v_2$, like $F(v_1, v_2)=\frac{f(v_1, v_2)}{\langle v_1, v_2\rangle }$, we roughly refer
$f(v_1, v_2)|_{v_1=\alpha_1 v_0, v_2=\alpha_2 v_0}$ as the `residue' of $F(v_1, v_2)$ at the pole where $\langle v_1, v_2\rangle=0$ and simply denote it
as ${\text{Res}}_{\lan v_1, v_2\ran=0} F(v_1, v_2)$.

\begin{figure}[t!]
\centering
\includegraphics[scale=0.5]{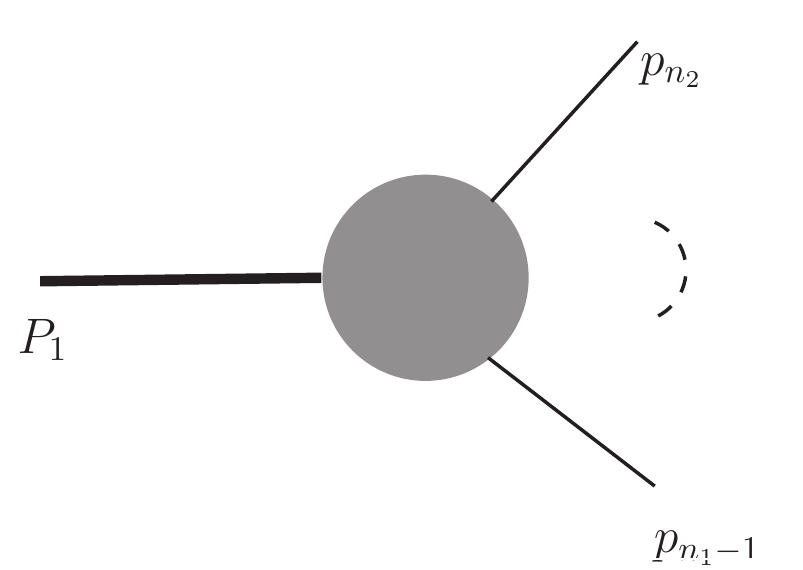}
\caption{One-off-shell-line sub-amplitude $A^1(P_1^{h_{P_1}}, n_2, \cdots, n_1-1)$.}
\label{fig:one-off-shell-line}
\end{figure}

\begin{figure}[t!]
\centering
\includegraphics[scale=0.5]{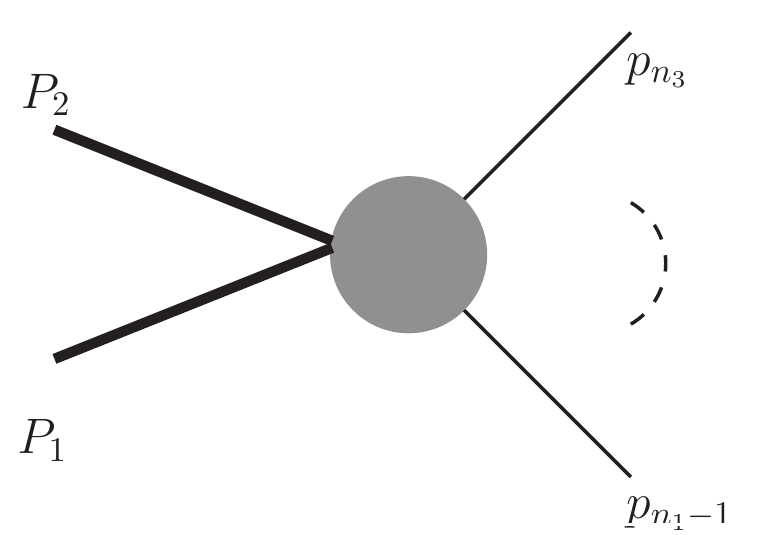}
\caption{Two-off-shell-line sub-amplitude $A^2(P_1^{h_{P_1}}, P_2^{h_{P_2}}, n_3, \cdots, n_1-1)$.}
\label{fig:two-off-shell-line}
\end{figure}

As in \cite{Zhu:2004kr}, we introduce two kinds of `off-shell amplitudes' constructed by CSW rules. We define $A^1(P_1^{h_{P_1}}, n_2, \cdots, n_1-1)$ to be the amplitude with one off-shell leg $P_1$, see Fig.~\ref{fig:one-off-shell-line}. The all-but-one-negative-helicity one-off-shell-line amplitudes were computed in \cite{Zhu:2004kr}. The other off-shell amplitude (see Fig.~\ref{fig:two-off-shell-line}), called $A^2(P_1^{h_{P_1}}, P_2^{h_{P_2}}, n_3, \cdots, n_1-1 )$, is defined as the amplitude constructed from all MHV diagrams with two adjacent off-shell legs $P_1, P_2$ directly attached to the same MHV vertex and the rest by the standard rules (with external momenta $P_{1,2}$ and $p_{n_3}, \cdots, p_{n_1-1}$).

A special note is in order. Because of the constraint that off-shell legs $P_1, P_2$ must be directly attached to the same MHV vertex, $A^2(P_1^{h_{P_1}}, P_2^{h_{P_2}}, n_3, \cdots, n_1-1 )$ is not a sub-amplitude in the conventional sense. They can be used to calculate amplitudes. They will be used in the analysis of spurious amplitude poles. However, it is sometimes referred to as a sub-amplitude by us. Also, because the helicities of the external gluons are fixed, we will not explicitly state them.

For generic external momenta,  the `residue' of $A^2(P_1^{h_{P_1}}, P_2^{h_{P_2}}, n_3, \cdots, n_1-1 )$ at the spurious pole at $\lan v_1, v_2\ran=0$ is the following,
\bea
&&\text{Res}_{\lan v_1, v_2 \ran=0}A^2(P_1^{+}, P_2^{+}, n_3, \cdots, n_1-1)=\frac{\alpha_3^2}{\alpha_1\alpha_2} A^1((P_1+P_2)^+, n_3, \cdots, n_1-1)\label{eq:plusplus}\,,\label{eq:res1}\\
&&\text{Res}_{\lan v_1, v_2 \ran=0}A^2(P_1^{+}, P_2^{-}, n_3, \cdots, n_1-1)=\frac{\alpha_2^3}{\alpha_1\alpha_3^2} A^1((P_1+P_2)^-, n_3, \cdots, n_1-1)\label{eq:plusminus}\,,\label{eq:res2}\\
&&\text{Res}_{\lan v_1, v_2 \ran=0}A^2(P_1^{-}, P_2^{+}, n_3, \cdots, n_1-1)=\frac{\alpha_1^3}{\alpha_2\alpha_3^2} A^1((P_1+P_2)^-, n_3, \cdots, n_1-1)\label{eq:minusplus}\,,\label{eq:res3}\\
&&\text{Res}_{\lan v_1, v_2 \ran=0}A^2(P_1^{-}, P_2^{-}, n_3, \cdots, n_1-1)=0\,.\label{eq:minusminus}
\eea

The proof of eq.~\eqref{eq:plusplus} is actually not very difficult. To get a feeling of what is involved in the proof, we refer the readers to Appendix A for an explicit example, $A^2(p_{1,2}^+, p_{3,4}^+, 5, 6, 7)$. $A^2(P_1^+, P_2^+, n_3, \cdots, n_1-1)$ is given by a summation over all needed diagrams. In each (connected) diagram there is an MHV vertex with two adjacent lines with $\lambda_{P_1}=v_1$ and $\lambda_{P_2}=v_2$. Because we assume the helicities of both $P_1$ and $P_2$ to be plus, the dependence on $v_{1,2}$ will only be:
 \be {1\over \langle \lambda_L, v_1\rangle \langle v_1,v_2\rangle
 \langle v_2, \lambda_R\rangle}\,,\label{eq:simple}
 \ee
where $\lambda_L$ is the spinor of the corresponding momentum just before  $P_1$, and $\lambda_R$ is the spinor of the corresponding momentum just after $P_2$, (here 'before' and 'after' are based on the cyclic order inside the MHV vertex involving $P_1$ and $P_2$). For $\langle v_1,v_2\rangle=0$ this gives a residue:
\be  {\alpha_3^2\over \alpha_1\alpha_2}\, {1\over \langle \lambda_L,v_3\rangle  \langle v_3,\lambda_R\rangle} .
\ee
A complete proof follows.

\begin{figure}[t!]
\centering
\includegraphics[scale=0.5]{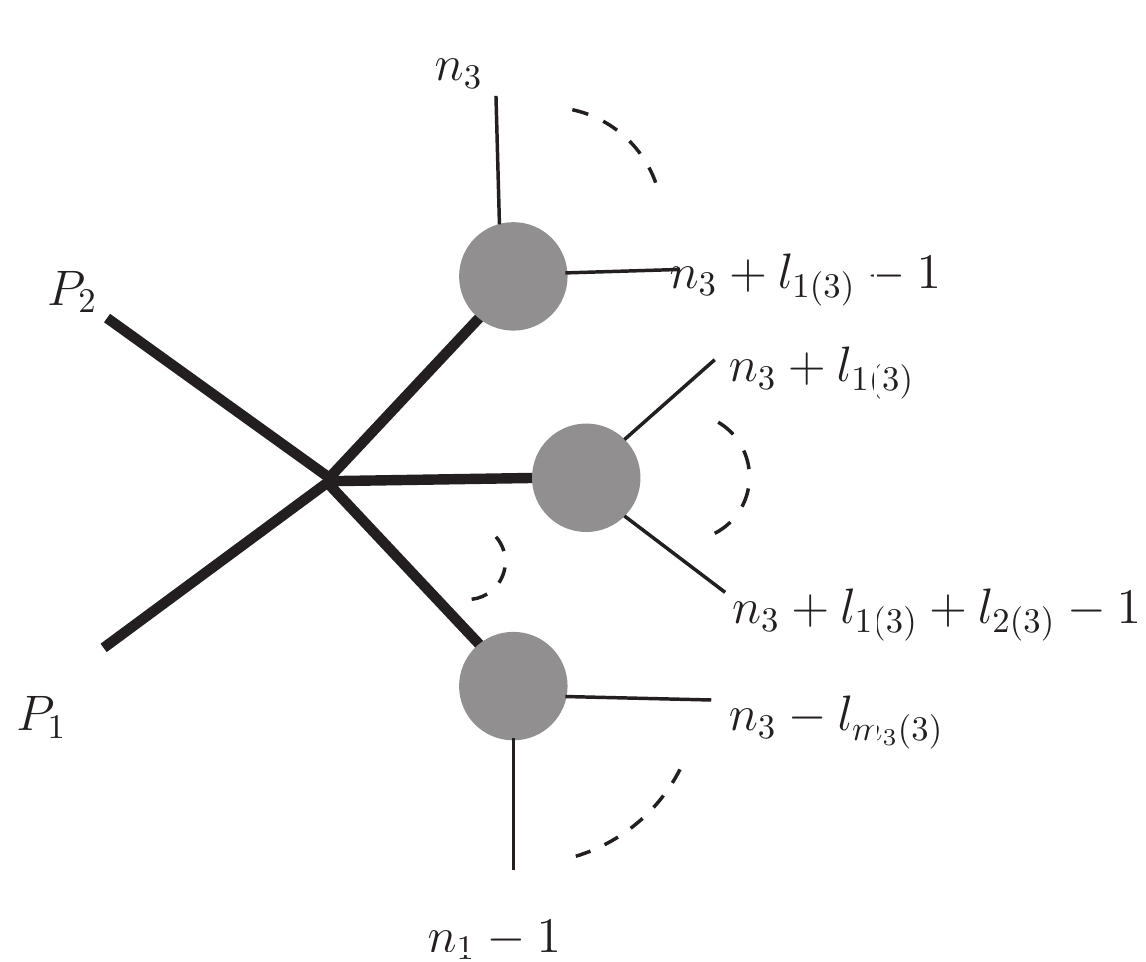}
\caption{Diagram decomposition for $ A^2(P_1^+, P_2^+, n_3, \cdots, n_1-1)$.}
\label{fig:diagram-decomposition}
\end{figure}

From the diagram decomposition in Fig.~\ref{fig:diagram-decomposition}, we have,
\bea
&& \hskip -2cm A^2(P_1^+, P_2^+, n_3, \cdots, n_1-1)=\sum_{m_3\geq 2}\, \sideset{}{^{\prime}}\sum_{l_{1(3)}, \cdots, l_{m_3(3)}} \sum_{h_{1(3)}, \cdots, h_{m_3(3)}}V_{m_3+2}(P_1^+, P_2^+, q_{1(3)}^{h_{q_{1(3)}}}, \cdots, q_{m_3(3)}^{h_{q_{m_3(3)}}}) \nonumber \\ &\times &A(n_3, \cdots, n_3+l_{1(3)}-1, -q_{1(3)}^{-{h_{{q_{1(3)}}}}})\frac{1}{q_{1(3)}^2}
A(n_3+l_{1(3)}, \cdots, n_3+l_{1(3)}+l_{2(3)}-1, -q_{2(3)}^{-{ h_{q_{2(3)}}}})\nonumber \\
&\times &\frac{1}{q_{2(3)}^2}\times \cdots \times A(n_1-l_{m_3(3)}, \cdots, n_1-1, -q_{m_3(3)}^{-{ h_{q_{m_3(3)}}}})\frac{1}{q_{m_3(3)}^2}\,,\eea
where the symbol $\sum^\prime$ in the second summation means a sum over $l_{1(3)}, \cdots, l_{m_3(3)}$ under the following constraints
\be l_{i(3)}>0, \,\, i=1, \cdots, m_3, \sum_{i=1}^{m_3}l_{i(3)}=n_1-n_3\,, \ee
and \be q_{i(3)}=p_{n_3+\sum_{j=1}^{i-1}l_{j(3)}, n_3+\sum_{j=1}^{i}l_{j(3)}-1} \ee
Here and below, the vertex like $V_{m_3+2}$ is understood to be vanishing if it does not satisfy the MHV conditions.

The  involved  MHV vertex $V_{m_3+2}(P_1^+, P_2^+, q_{1(3)}^{h_{q_{1(3)}}}, \cdots, q_{m_3(3)}^{h_{q_{m_3(3)}}})$ is,
\bea && \hskip -2cm V_{m_3+2}(P_1^+, P_2^+, q_{1(3)}^{h_{q_{1(3)}}}, \cdots, q_{m_3(3)}^{h_{q_{m_3(3)}}})\nonumber\\
&=&\frac{\langle \tilde{v}_{j(3)}, \tilde{v}_{k(3)} \rangle^4}{\langle v_1, v_2\rangle \langle v_2, \tilde{v}_{1(3)}\rangle \,( \prod_{i=1}^{m_3-1}
\langle \tilde{v}_{i(3)}, \tilde{v}_{(i+1)(3)} \rangle) \, \langle \tilde{v}_{m_3(3)}, v_1\rangle}\,,\label{eq:vertex} \eea
with $m_3\geq 2$. Here $\tilde{v}_{i(3)}$ is the spinor corresponding to the momentum $q_{i(3)}$ based on the CSW rules, and we have assumed
\be h_{q_{j(3)}}=h_{q_{k(3)}}=-1, \,\, h_{q_{i(3)}}=+1, \text{ for}  \, 1\leq i\leq m_3, i\neq j, k.  \ee
Notice that $\tilde{v}_{m_3(3)}$ and $\tilde{v}_{1(3)}$ in \eqref{eq:vertex} are just $\lambda_L$ and $\lambda_R$ in \eqref{eq:simple}, respectively.
The `residue' of this vertex at the pole $\langle v_1, v_2 \rangle=0$ is
\bea && \hskip -2cm {\text{Res}}_{\lan v_1, v_2\ran=0}V_{m_3+2}(P_1^+, P_2^+, q_{1(3)}^{h_{q_{1(3)}}}, \cdots, q_{m_3(3)}^{h_{q_{m_3(3)}}})\nonumber\\
&=&\frac{\alpha_3^2}{\alpha_1 \alpha_2}  \frac{\langle \tilde{v}_{j(3)}, \tilde{v}_{k(3)} \rangle^4}{ \langle -v_3, \tilde{v}_{1(3)}\rangle \,(\prod_{i=1}^{m_3-1}
\langle \tilde{v}_{i(3)}, \tilde{v}_{(i+1)(3)} \rangle) \,\langle \tilde{v}_{m_3(3)}, -v_3\rangle}\nonumber\\
&=&\frac{\alpha_3^2}{\alpha_1\alpha_2} V_{m_3+1}(-P_3^+,  q_{1(3)}^{h_{q_{1(3)}}}, \cdots, q_{m_3(3)}^{h_{q_{m_3(3)}}})\,.
\label{eq:residue1}\eea
We can see that the `residue' of MHV vertex at $\lan v_1, v_2 \ran=0 $ is proportional to another MHV vertex with the number of legs reduced by one. 
This simple fact inherited from the collinear limit of MHV amplitudes plays a crucial role in our proof.
So the residue of $A^2(P_1^+, P_2^+, n_3, \cdots, n_1-1)$ at the same pole is
\bea && \hskip -2cm \text{Res}_{\lan v_1, v_2\ran=0}A^2(P_1^+, P_2^+, n_3, \cdots, n_1-1)=\frac{\alpha_3^2}{\alpha_1 \alpha_2}\sum_{m_3\geq 2}\,\, \sideset{}{^\prime}\sum_{l_{1(3)}, \cdots, l_{m_3(3)}} \sum_{h_{1(3)}, \cdots, h_{m_3(3)}}\nonumber\\
&& V_{m_3+1}(-P_3^+, q_{1(3)}^{h_{q_{1(3)}}}, \cdots, q_{m_3(3)}^{h_{q_{m_3(3)}}}) A(n_3, \cdots, n_3+l_{1(3)}-1, -q_{1(3)}^{-{ h_{q_{1(3)}}}})\frac{1}{q_{1(3)}^2}
\nonumber \\ &\times &A(n_3+l_{1(3)}, \cdots, n_3+l_{1(3)}+l_{2(3)}-1, -q_{2(3)}^{-{ h_{q_{2(3)}}}})\frac{1}{q_{2(3)}^2} \cdots \nonumber\\
&\times& A(n_1-l_{m_3(3)}, \cdots, n_1-1, -q_{m_3(3)}^{-{  h_{q_{m_3(3)}}}})\frac{1}{q_{m_3(3)}^2}\,.\eea

Notice that we have the following result
\bea \label{eq:resummation1} &&\hskip -2cm \sum_{m_3\geq 2}\,\, \sideset{}{^\prime}\sum_{l_{1(3)}, \cdots, l_{m_3(3)}} \sum_{h_{1(3)}, \cdots, h_{m_3(3)}}V_{m_3+1}(-P_3^+, q_{1(3)}^{h_{q_{1(3)}}}, \cdots, q_{m_3(3)}^{h_{q_{m_3(3)}}}) \nonumber \\ &\times &A(n_3, \cdots, n_3+l_{1(3)}-1, -q_{1(3)}^{-{ h_{q_{1(3)}}}})\frac{1}{q_{1(3)}^2}\nonumber\\
&\times&
A(n_3+l_{1(3)}, \cdots, n_3+l_{1(3)}+l_{2(3)}-1, -q_{2(3)}^{-{ h_{q_{2(3)}}}})\nonumber \\
&\times &\frac{1}{q_{2(3)}^2}\times \cdots \times A(n_1-l_{m_3(3)}, \cdots, n_1-1, -q_{m_3(3)}^{-{ h_{m_3(3)}}})\frac{1}{q_{m_3(3)}^2}\nonumber \\
&=& A^1(n_3, \cdots, n_1-1, -P_3^+)\,, \eea
from the diagram decomposition for the amplitude $A^1(n_3, \cdots, n_1-1, -p_{n_3, n_1-1}^+)$ which is the same as Fig.~\ref{fig:diagram-decomposition} but with the two off-shell momentum lines $P_1$ and $P_2$ combined into one momentum line
$-P_3=P_1+P_2$.\footnote{{ Similar decomposition was used in \cite{Wu:2004fba} to demonstrate that the amplitudes from CSW rules satisfying the charge conjugation identity
(also known as the color-ordered reversed relation) and the dual Ward identity (also known as the  $U(1)$ decoupling relation).}}
This conclude the proof of eq.~\eqref{eq:plusplus}.

The proofs of other two identities, eq.~\eqref{eq:plusminus} and eq.~\eqref{eq:minusplus}, are nearly identical. The only point to note is that there is an extra factor $\langle v_{1,2},\lambda_*\rangle^4$ from the MHV amplitude because one of the off-shell lines has negative helicity. ($\lambda_*$ is the spinor of the other negative helicity line.) This only changes the residue's prefactor.
The proof of the last identity, eq.~\eqref{eq:minusminus}, is trivial because there is an extra factor $\langle v_{1},v_2\rangle^4$ in the numerator and there is no pole.

\subsection{ Lorentz covariance of $A_{\text{CSW}}$\label{subsection:cancellation}}

In this section, we will show that $A_{\text{CSW}}$ is independent of $\tilde\eta$. In other words, the resulting amplitude is Lorentz covariant.

It is not difficult to see that $A_{\text{CSW}}$ is invariant under the transformation $\tilde{\eta}^{\dot{\alpha}}\to t\tilde{\eta}^{\dot{\alpha}}$. This is derived from the fact that the MHV amplitude (as general amplitudes) is scaled by a factor $t_i^{-2h_i}$ when we scale $\lambda_i$ to $t_i\lambda_i$ for fixed $i$ and the two ends of each propagator in MHV diagrams have opposite helicities.

Then if we define $\tilde{t}={\tilde{\eta}^1}/{\tilde{\eta}^2}$, $A_{\text{CSW}}$ can be considered as a function of $\lambda_i, \tilde{\lambda}_i$, and $\tilde{t}$. We will show that  $A_{\text{CSW}}$ is a holomorphic function of $\tilde{t}\in \hat{{\mathbf C}}(\equiv{\mathbf C} \cup \{\infty\})$ for generic external momenta, which leads to the result that $A_{\text{CSW}}$ is independent of $\tilde{\eta}$. To do this, we simply need to demonstrate that the residue for any arbitrary $\tilde{t}$ disappear.

Let us consider the non-degenerate spurious poles mentioned previously and demonstrate that all such poles of ${\tilde t}$ cancel among themselves. The proof of the cancelation of spurious pole for the degenerate case is very similar to the non-degenerate case, and we will briefly discuss these cases in Appendix~\ref{appendix:degenerate}.

\begin{figure}[t!]
\centering
\includegraphics[scale=0.5]{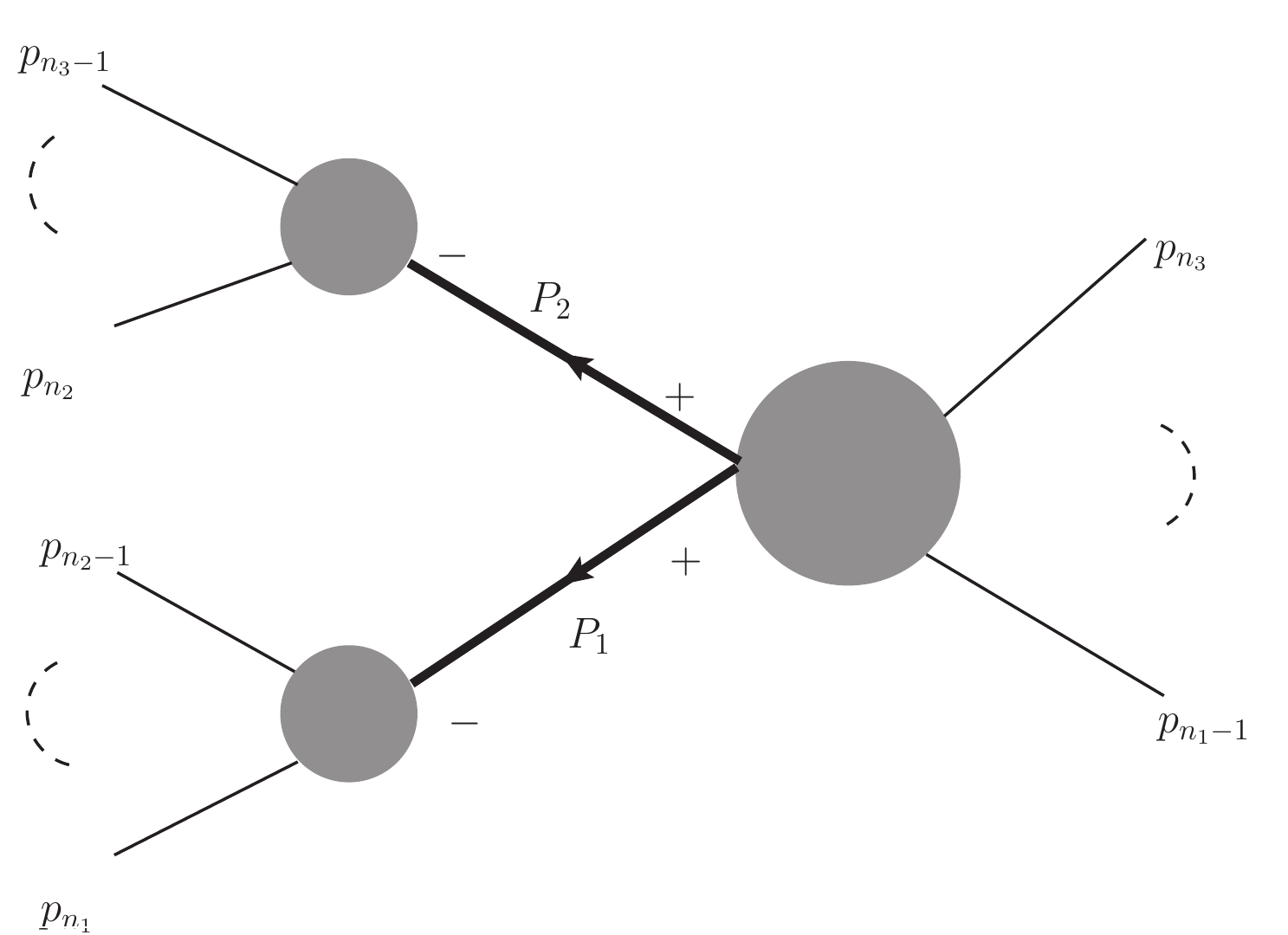}
\caption{Diagram contributing to $R_3$.}
\label{fig:fig3}
\end{figure}

Consider the poles from $\langle v_1, v_2\rangle=0$ with $h_{P_1}=h_{P_2}=+1$, we will show these poles will be canceled by poles from $\langle v_2, v_3\rangle=0$  with  $h_{P_2}=+1, h_{P_3}=-1$, and the poles from $\langle v_3, v_1\rangle=0$ with $ h_{P_3}=-1, h_{P_1}=+1$.

The  pole from $\langle v_1, v_2 \rangle=0$ (recall that we have assumed $h_{P_1}=h_{P_2}=+1$) is from the following contributions (see Fig.~\ref{fig:fig3}),
\be R_3=A^1(-P_1^-, n_1, \cdots, n_2-1)\frac{1}{P_1^2}A^1( -P_2^-, n_2, \cdots, n_3-1)\frac1{P_2^2}A^2(P_1^+, P_2^+, n_3, \cdots, n_1-1)\,.
\ee
By using eq.~\eqref{eq:res1}, the `residue' of these contributions at $\lan v_1, v_2\ran=0$ is
\bea\tilde{R}_3&=& \frac{\alpha_3^2}{\alpha_1\alpha_2}A^1( -P_1^-, n_1, \cdots, n_2-1)\frac{1}{P_1^2}A^1( -P_2^-, n_2, \cdots, n_3-1)\nonumber\\
&\times&\frac1{P_2^2}A^1( -P_3^+, n_3, \cdots, n_1-1).\eea

We now consider the pole at $\lan v_2, v_3\ran=0$ \footnote{Recall that $\lan v_1, v_2 \ran=\lan v_2, v_3\ran=\lan v_3, v_1 \ran$ even away from the poles.} with $h_{P_2}=+1, h_{P_3}=-1$. Such pole is from
 \be R_1= A^1(-P_2^-, n_2, \cdots, n_{3}-1) \frac{1}{P_2^2} A^1(-P_3^+, n_3, \cdots, n_1-1) \frac{1}{P_3^2} A^2(P_2^+, P_3^-, n_1, \cdots, n_2-1)\,.
 \ee
Using eq.~\eqref{eq:res2}, the residue of the above contributions at  $\lan v_2, v_3\ran(=\lan v_1, v_2 \ran)=0$ is
\bea
\tilde{R}_1
&=&\frac{\alpha_3^3}{\alpha_1^2 \alpha_2} A^1(-P_2^-, n_2, \cdots, n_{3}-1) \frac{1}{P_2^2}\nonumber \\
&\times & A^1(-P_3^+, n_3, \cdots, n_1-1) \frac{1}{P_3^2} A^1(-P_1^-, n_1, \cdots, n_2-1)\,,
\eea

Now we turn to  consider the pole at $\lan v_3, v_1\ran=0$ with  $h_{P_3}=-1, h_{P_1}=+1$ from
\be R_2 = A^1(-P_3^+, n_3, \cdots, n_1-1)\frac{1}{P_3^2}A^1(-P_1^-, n_1, \cdots, n_2-1) \frac1{P_1^2}A^2(P_3^-, P_1^+, n_2, \cdots, n_3-1)\,.
 \ee
From eq.~\eqref{eq:res3}, we know that the residue of the above contributions at $\lan v_3, v_1 \ran (=\lan v_1, v_2 \ran)=0$ is
\bea \tilde{R}_2&=& \frac{\alpha^3}{\alpha_1 \alpha_2^2}A^1(-P_3^+, n_3, \cdots, n_1-1)\frac{1}{P_3^2}A^1(-P_1^-, n_1, \cdots, n_2-1) \frac1{P_1^2}\nonumber \\
&\times& A^1(-P_2^-, n_2, \cdots, n_3-1)\,.
 \eea
 The sum of $\tilde{R}_1, \tilde{R}_2, \tilde{R}_3$ is
 \bea\sum_{i=1}^3\tilde{R}_i &=& \left(\frac{P_1^2}{\alpha_1}+\frac{P_2^2}{\alpha_2}+\frac{P_3^2}{\alpha_3}\right)\frac{\alpha_3^3}{\alpha_1 \alpha_2} A^1(-P_1^-, n_1, \cdots, n_2-1)\frac1{P_1^2}\nonumber \\
 &\times&A^1(-P_2^-, n_2, \cdots, n_3-1)\frac1{P_2^2}A^1(-P_3^+, n_3, \cdots, n_1-1)\frac1{P_3^2}\,. \eea

 We will prove in Appendix~\ref{appendix:identity} that
 \be\frac{P_1^2}{\alpha_1}+\frac{P_2^2}{\alpha_2}+\frac{P_3^2}{\alpha_3}=0\,,\label{eq:identity} \ee
 and this leads to
 \be \sum_{i=1}^3 \tilde{R}_i=0\,.\ee

 Notice that there are other poles from $\lan v_1, v_2\ran=0$, but these poles can be  divided into the following two groups:
\begin{itemize}
\item  poles from $\langle v_1, v_2\rangle=0$ with $h_{P_1}=+1, h_{P_2}=-1$,  poles from $\langle v_2, v_3\rangle=0$  with  $h_{P_2}=-1, h_{P_3}=+1$, and poles from $\langle v_3, v_1\rangle=0$ with $ h_{P_3}=h_{P_1}=+1$;

\item  poles from $\langle v_1, v_2\rangle=0$ with $h_{P_1}=-1, h_{P_2}=+1$,  poles from $\langle v_2, v_3\rangle=0$  with  $h_{P_2}=h_{P_3}=+1$, and  poles from $\langle v_3, v_1\rangle=0$ with $ h_{P_3}=+1, h_{P_1}=-1$.

\end{itemize}

One can demonstrate that the poles in each group cancel each other, just as we did above.

The above classification of non-degenerate poles at $\langle v_1, v_2\rangle(=\langle v_2, v_3\rangle=\langle v_3, v_1\rangle)=0 $ can be summarized into  Table~\ref{table1}.

As previously stated, the helicity configuration residues in each line cancel each other out.

\begin{table}
  \centering
  \caption{Classification of non-degenerate poles.}\label{table1}
\begin{tabular}{|c|c|c|c|}
\hline
poles & $\langle v_1, v_2\rangle=0$ & $\langle v_2, v_3\rangle=0$  & $\langle v_3, v_1\rangle=0$ \\
\hline
      &$h_{P_1}=h_{P_2}=+1$ & $h_{P_2}=+1, h_{P_3}=-1$&$h_{P_3}=-1, h_{P_1}=+1$\\
\hline

      &$h_{P_1}=+1, h_{P_2}=-1$ & $h_{P_2}=-1, h_{P_3}=+1$&$h_{P_3}=h_{P_1}=+1$\\
\hline
&$h_{P_1}=-1, h_{P_2}=+1$ & $h_{P_2}=h_{P_3}=+1$&$h_{P_3}=+1, h_{P_1}=-1$\\
\hline
\end{tabular}
\end{table}

\subsection{ Completion of the proof\label{subsection:completion}}
This part of the proof is not new.  The ingredients have appeared in \cite{Cachazo:2004kj} and \cite{Britto:2005fq}. After proving that  $A_{\text{CSW}}$ is independent of $\tilde{\eta}$, we only need to consider  poles of $A_{\text{CSW}}$ as a function of $\lambda, \tilde{\lambda}$. In \cite{Cachazo:2004kj}, it was shown that the CSW rules correctly reproduce all the collinear singularities and multi-gluon singularities. So $A_{\text{CSW}}-A_{\text{Feynman}}$ can only be a polynomial of $\lambda_i, \tilde{\lambda}_i$ as pointed in \cite{Britto:2005fq}.
In \cite{Britto:2005fq}, as a step to their proof of CSW rules, it was noticed that the mass dimension of tree amplitudes with $n$ gluon is $4-n$. We will explicitly show that both $A_{\text{CSW}}$ and $A_{\text{Feynman}}$ have mass dimension $4-n$ in Appendix~\ref{appendix:dimensions}.
This leads to $A_{\text{CSW}}=A_{\text{Feynman}}$ when $n>4$.
Also as mentioned in \cite{Britto:2005fq}, the case with $n=4$ can be confirmed directly.
This completes the proof.

\section{Conclusion\label{section:conclusion}}

In this paper, we presented a new proof of the CSW rules for gluonic tree amplitudes using MHV diagrams. Our new contribution provides explicit evidence for the cancellation of spurious poles caused by the off-shell continuation of MHV amplitudes. 
This leads to the conclusion that $A_{\text{CSW}}$ is Lorentz invariant. This step is based on an examination of spurious poles of specially-defined two-off-shell-line sub-amplitudes, and we discover that the `residue' is proportional to certain one-off-shell-line sub-amplitudes. 

It would be interesting to investigate the applicability of our method to other theories or loop levels.
Our new approach can be applied to amplitudes from CSW rules involving fermions \cite{hep-th/0404072} by similarly generalizing the treatment in \cite{hep-th/0406146}. It is interesting to study the generalization of our approach to CSW rules for QED \cite{hep-th/0509063} and theories involving massive particles like Higgs bosons \cite{hep-th/0411092, hep-ph/0412167, hep-th/0412275, hep-th/0504159, hep-th/0507292}.  The situation with gravity amplitudes is much more complicated. Risager's idea \cite{hep-th/0508206} was used in \cite{hep-th/0509016} to propose CSW-like rules for tree-level graviton amplitudes\footnote{Earlier attempts can be found in \cite{hep-th/0405086, Wu:2004fba}.}. However, it was discovered that this proposal only works when the number of gravitons $n$ is less than $12$ \cite{0805.0757, Benincasa:2007qj}; see also further studies in \cite{1205.3500, 1212.6257}. It is interesting to use our approach to demonstrate that the spurious poles are cancelled among themselves when $n<12$. This cancellation fails when $n\ge 12$. We hope that these studies provide hints on how to develop CSW-like rules that are applicable in all cases.

CSW rules were successfully generalized to the one-loop level by reproducing some amplitudes in ${\cal N}=4$ SYM theories \cite{hep-th/0407214, hep-th/0410280, hep-th/0410278, hep-th/0410045, hep-th/0410118}. It is interesting to use our approach to determine whether the integrand of the planar one-loop amplitudes from CSW rules \cite{hep-th/0407214, hep-th/0410280, hep-th/0410278} has no spurious poles or these poles only disappear after loop integration.

\section*{Acknowledgments}
Jun-Bao Wu would like to thank Bin Chen, Jian-Xin Lu, Gang Yang for their very helpful discussions. He would also like to thank Peking University since a key idea was obtained when he walked crossing the street between the east gate and the School of Physics. Chuan-Jie Zhu would like to thank Bo Feng for the discussions. We would also like to thank Zhengwen Liu for very helpful suggestions after reading the manuscript.
The work of Jun-Bao Wu  and Wen-Jie Zhang was supported by the National Natural Science Foundation of China, Grant No.~11975164, 11935009, 12047502, 11947301,
and  Natural Science Foundation of Tianjin under Grant No.~20JCYBJC00910.  The work of Chuan-Jie Zhu was supported in part by a fund from Hunan University of Arts and Science. The unusual ordering of authors instead of the standard alphabetical one in the hep-th community is for students to get proper recognition of contribution under the current practice in China.
  The diagrams were drawn using the JavaDraw package \cite{Binosi:2003yf, Binosi:2008ig}.

\appendix

\section{Cancelation of non-degenerate spurious poles: an example}
For the convenience of the readers, we use the googly amplitude $A(1^+, 2^-, 3^+, 4^-, 5^-, 6^-, 7^-)$ to display the procedure of
the cancelation of non-degenerate spurious poles. We consider the situation with $n_1=1, n_2=3, n_3=5$.

\begin{figure}[t!]
\centering
\includegraphics[scale=0.5]{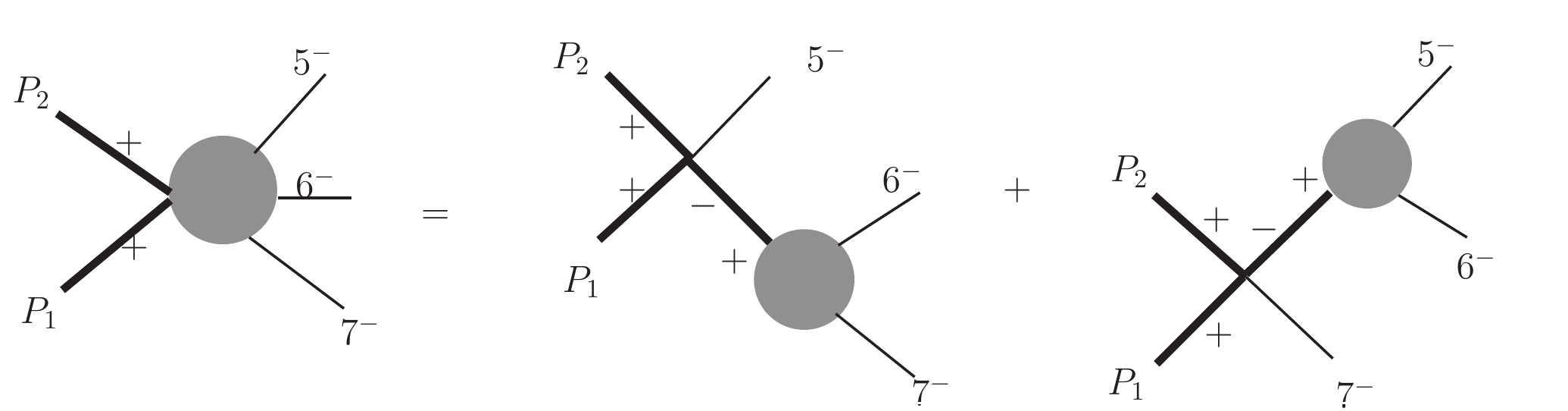}
\caption{Diagrammatic decomposition for $A^2(p_{1, 2}^+, p_{3, 4}^+, 5, 6, 7)$.}
\label{fig:eq48}
\end{figure}

The diagrams   with a pole at   ${\lan \lambda_{p_{1, 2}}, \lambda_{p_{3, 4}}\ran}=0$ and satisfying $h_{p_{1, 2}}=h_{p_{3, 4}}=+1$  give contributions proportional to  the two-off-line sub-amplitude $A^2(p_{1,2}^+, p_{3,4}^+, 5, 6, 7)$:
 \be
 R_3=A^1(1, 2, -p_{1,2}^-)\, {1\over p_{1,2}^2}\,
  A^1(3, 4,-p_{3,4}^-)\, {1\over p_{3,4}^2}\,A^2(p_{1,2}^+, p_{3,4}^+, 5, 6, 7)\,.
  \ee
where
\bea
 A^1(1, 2, -p_{1,2}^-) &=& \frac{\lan \lambda_2, \lambda_{p_{1, 2}}\ran^3}{\lan 1, 2 \ran \lan \lam_{p_{1, 2}}, \lam_1 \ran}\,,\\
 A^1(3, 4, -p_{3,4}^-)&=&\frac{\lan\lam_4, \lam_{p_{3, 4}} \ran^3}{\lan 3, 4 \ran \lan \lam_{p_{3, 4}}, \lam_3\ran}\,,\\
A^2(p_{1,2}^+, p_{3,4}^+,5, 6, 7) &=&
\frac{ \lan \lam_5, \lam_{p_{6, 7}} \ran^3 }{\lan \lam_{p_{6, 7}}, \lam_{p_{1, 2}}\ran \lan \lam_{p_{1, 2}}, \lam_{p_{3, 4}} \ran \lan \lam_{p_{3, 4}}, \lam_5 \ran }\frac{1}{p_{6, 7}^2} A^1(-p_{6,7}^+,p_6^-,p_7^-) \nonumber\\
&+&\frac{\lan \lam_{p_{5, 6}}, \lam_7 \ran^3 }{\lan \lam_7, \lam_{p_{1, 2}} \ran \lan \lam_{p_{1, 2}}, \lam_{p_{3, 4}} \ran
\lan\lam_{p_{3, 4}}, \lam_{p_{5, 6}}  \ran }\frac{1}{p_{5, 6}^2} A^1(-p_{5,6}^+,p_5^-,p_6^-)\,,\\
A^1(-p_{6,7}^+, 6, 7)&=& \frac{\lan 6, 7 \ran^3}{\lan \lam_7, \lam_{p_{6, 7}} \ran \lan \lam_{p_{6, 7}}, \lam_6 \ran }\,,\\
A^1(-p_{5,6}^+, 5, 6) &=&
\frac{\lan 5, 6  \ran^3}{\lan\lam_6, \lam_{p_{5, 6}} \ran \lan\lam_{p_{5, 6}}, \lam_5 \ran}\,.
\eea
Here we have used the diagram decomposition for $A^2(p_{1,2}^+, p_{3,4}^+, 5, 6, 7)$ as shown in Fig.~\ref{fig:eq48}.

The `residue' of $A^2(p_{1,2}^+, p_{3,4}^+,5, 6, 7)$ at $\lan \lam_{p_{1, 2}}, \lam_{p_{3, 4}} \ran=0 $ is
\bea & & \hskip -2cm \frac{\alpha_3^2}{\alpha_1\alpha_2}\,\left(
\frac{ \lan \lam_5, \lam_{p_{6, 7}} \ran^3 }{\lan \lam_{p_{6, 7}}, \lam_{-p_{5, 7}}\ran \lan \lam_{-p_{5, 7}}, \lam_5 \ran } \frac{1}{p_{6, 7}^2} A^1(-p_{6,7}^+, 6, 7)  \right.\nonumber\\
& &  +\left.\frac{\lan \lam_{p_{5, 6}}, \lam_7 \ran^3 }{\lan \lam_7, \lam_{-p_{5, 7}} \ran
\lan\lam_{-p_{5, 7}}, \lam_{p_{5, 6}}  \ran }
\frac{1}{p_{5, 6}^2} A^1(-p_{5,6}^+, 5, 6)\right) \nonumber\\
&=& \frac{\alpha_3^2}{\alpha_1\alpha_2} \, A^1(-p_{5, 7}^+, 5, 6, 7)\,.
 \eea
 This gives the complete `residues' from $R_3$:
\be
\tilde{R}_3= \frac{\alpha_3^2}{\alpha_1\alpha_2} A^1(-p_{1, 2}^-, 1, 2) \frac1{p_{1, 2}^2} A^1(-p_{3, 4}^-, 3, 4)\frac1{p_{3, 4}^2}A^1(-p_{5, 7}^+, 5, 6, 7)\,.
\ee

\begin{figure}[t!]
\centering
\includegraphics[scale=0.5]{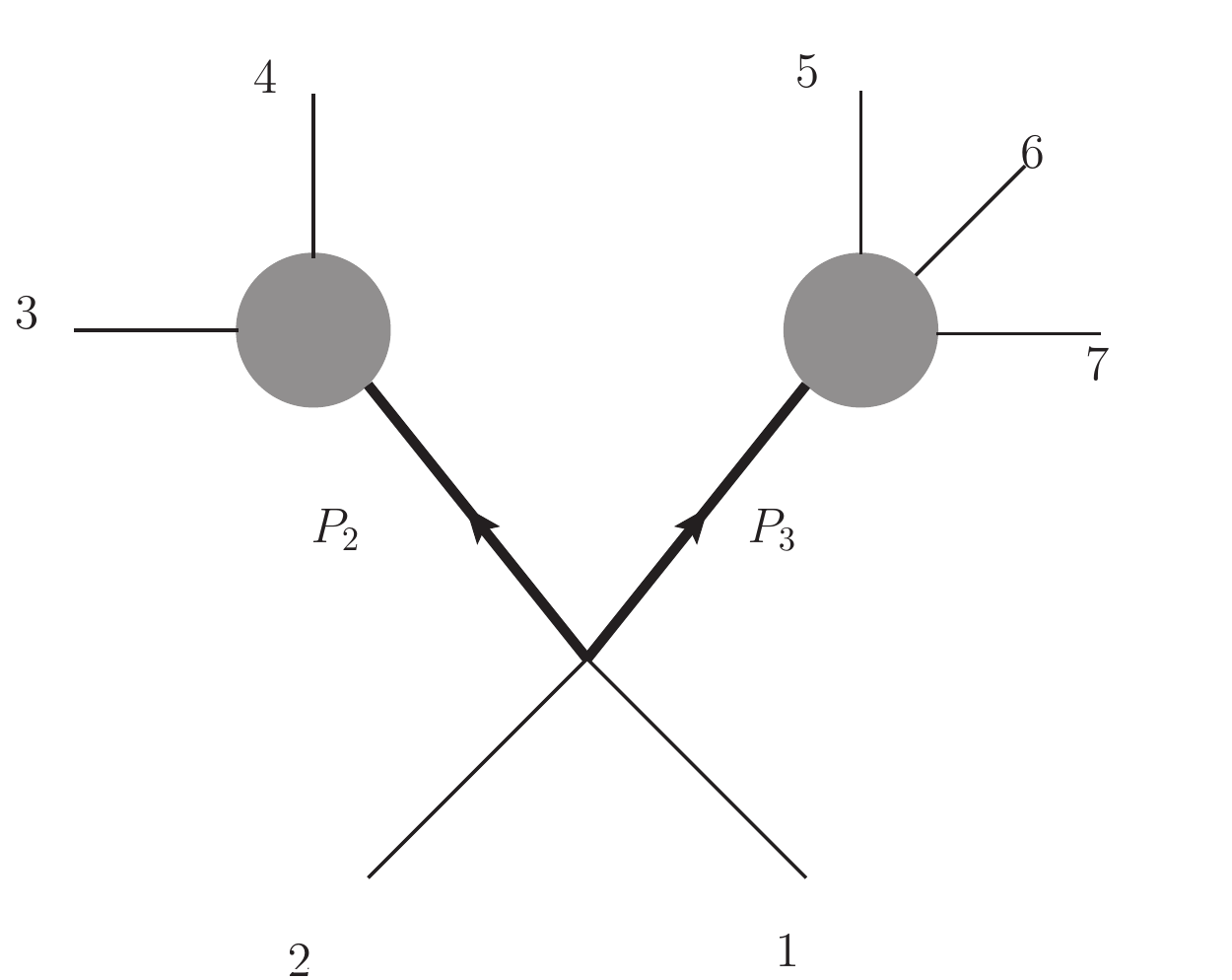}
\caption{Diagram giving contribution $R_1$ that has a pole at $\lan\lam_{p_{3, 4}}, \lam_{p_{5, 7}} \ran=0$.}
\label{fig:A2}
\end{figure}

\begin{figure}[t!]
\centering
\includegraphics[scale=0.5]{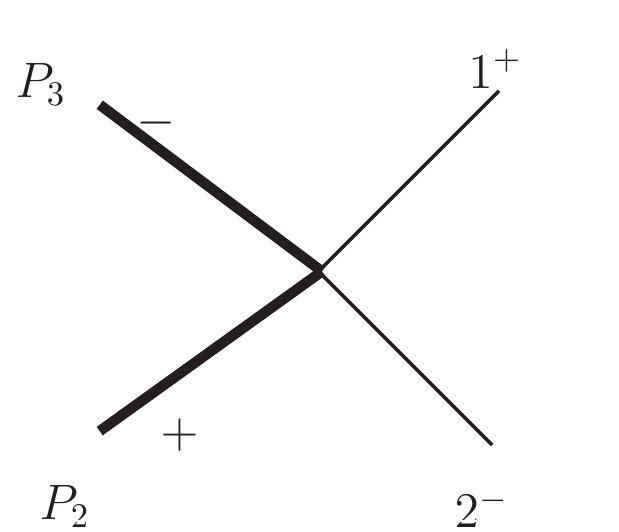}
\caption{Diagram contributing to the two-off-shell-line sub-amplitude $A^2(p_{3, 4}^+, p_{5, 7}^-, 1, 2)$.}
\label{fig:A23}
\end{figure}

The diagram involving a factor $1/\lan\lam_{p_{3, 4}}, \lam_{p_{5, 7}} \ran$ with $h_{p_{3, 4}}=+1, h_{p_{5, 7}}=-1$ is shown in Fig.~\ref{fig:A2}.
There is  only one diagram contributing to two-off-shell-line sub-amplitude $A^2(p_{3, 4}^+, p_{5, 7}^-, 1, 2)$, as shown in Fig.~\ref{fig:A23}.
This gives a contribution:
\bea R_1 &=& A^2(p_{3, 4}^+, p_{5, 7}^-, 1, 2)\frac1{p_{3, 4}^2} A^1(-p_{3, 4}^-, 3, 4) \frac1{p_{5, 7}^2} A^1(-p_{5, 7}^+, 5, 6, 7)  \nonumber\\
&=& \frac{\lan\lam_{p_{5, 7}}, \lam_2 \ran^4}{\lan 1, 2 \ran  \lan\lam_2, \lam_{p_{3, 4}} \ran  \lan\lam_{p_{3, 4}}, \lam_{p_{5, 7}} \ran \lan\lam_{p_{5, 7}}, \lam_1 \ran }\frac1{p_{3, 4}^2} A^1(-p_{3, 4}^-, 3, 4) \frac1{p_{5, 7}^2} A^1(-p_{5, 7}^+, 5, 6, 7) \,. \eea
The `residue' of $R_2$ at $\lan\lam_{p_{3, 4}}, \lam_{p_{5, 7}} \ran(=\lan \lam_{p_{1, 2}}, \lam_{p_{3, 4}} \ran)=0$ is
\bea \tilde{R}_1&=&\frac{\alpha_3^3}{\alpha_1^2 \alpha_2}\frac{\lan\lam_{-p_{1, 2}}, \lam_2 \ran^3}{\lan 1, 2\ran   \lan\lam_{-p_{1, 2}}, \lam_1 \ran }\frac1{p_{3, 4}^2} A^1(-p_{3, 4}^-, 3, 4) \frac1{p_{5, 7}^2} A^1(-p_{5, 7}^+, 5, 6, 7)\nonumber\\
&=&\frac{\alpha_3^3}{\alpha_1^2 \alpha_2} A^1(-p_{1, 2}^-, 1, 2)\frac1{p_{3, 4}^2} A^1(-p_{3, 4}^-, 3, 4) \frac1{p_{5, 7}^2} A^1(-p_{5, 7}^+, 5, 6, 7)\,. \eea

\begin{figure}[t!]
\centering
\includegraphics[scale=0.5]{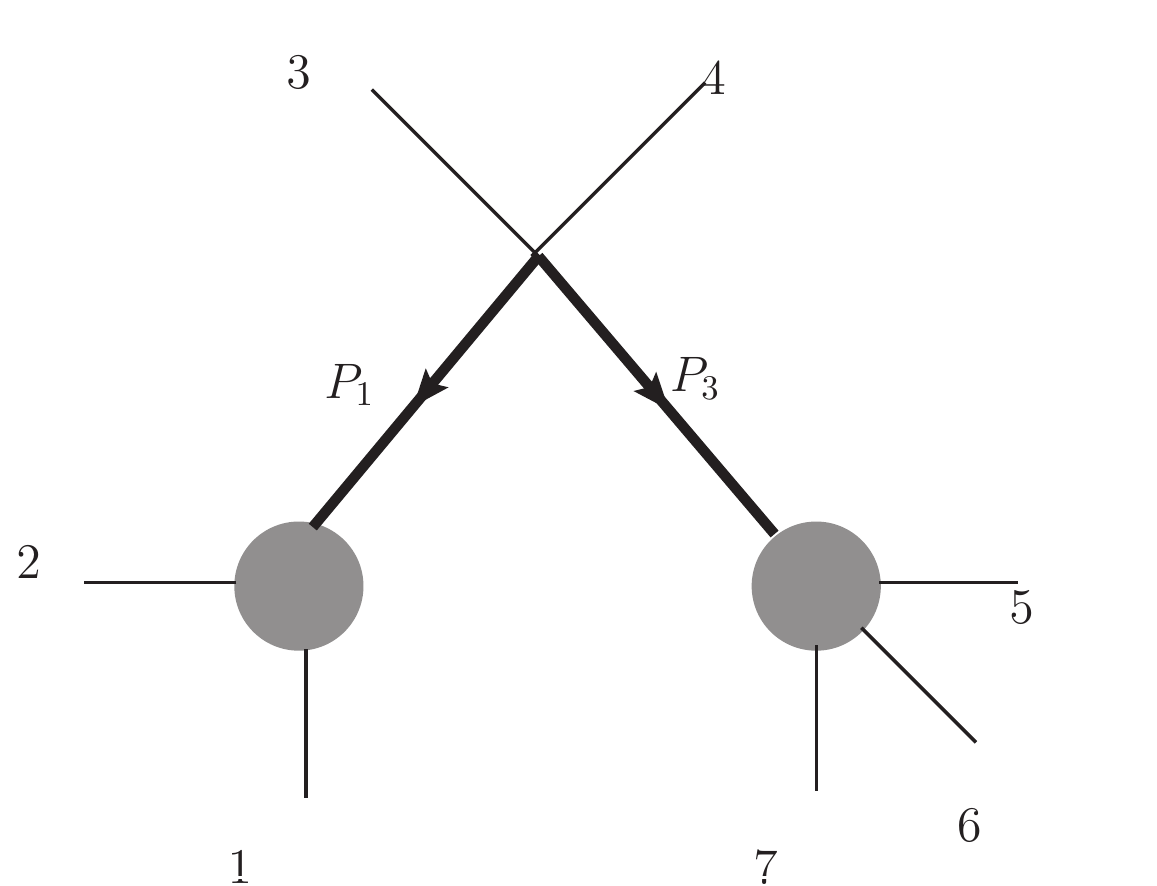}
\caption{Diagram giving contribution $R_2$ that has a pole at $\lan\lam_{p_{5, 7}}, \lam_{p_{1, 2}} \ran=0$.}
\label{fig:A22}
\end{figure}

\begin{figure}[t!]
\centering
\includegraphics[scale=0.5]{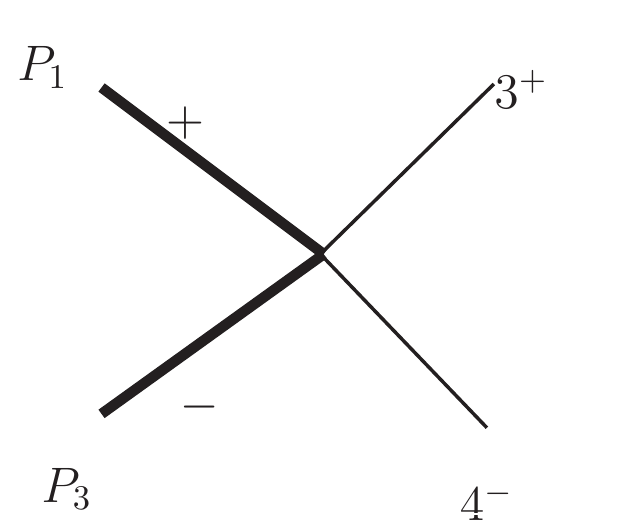}
\caption{Diagram contributing to the two-off-shell-line sub-amplitude $A^2(p_{5, 7}^-, p_{1, 2}^+, 3, 4)$.}
\label{fig:A24}
\end{figure}

The diagram involving pole at $\lan \lam_{p_{5, 7}}, \lam_{p_{1, 2}} \ran =0$  with $h_{p_{5, 7}}=-1, h_{p_{1, 2}}=+1$ is shown in Fig.~\ref{fig:A22}. The related two-off-shell-line sub-amplitude appearing here is  $A^2(p_{5, 7}^-, p_{1, 2}^+, 3, 4)$, as shown in Fig.~\ref{fig:A24}. This gives a contribution:
\bea R_2
 &=&A^1(-p_{1, 2}^-, 1, 2)\frac{1}{p_{1, 2}^2} A^1(-p_{5, 7}^+, 5, 6, 7) \frac{1}{p_{5, 7}^2}A^2(p_{5, 7}^-, p_{1, 2}^+, 3, 4) \nonumber\\
 &=& A^1(-p_{1, 2}^-, 1, 2)\frac{1}{p_{1, 2}^2} A^1(-p_{5, 7}^+, 5, 6, 7) \frac{1}{p_{5, 7}^2}\frac{\lan\lam_4, \lam_{p_{5, 7}} \ran^3 }{\lan \lam_{p_{5, 7}}, \lam_{p_{1, 2}}\ran\lan \lam_{p_{1, 2}, \lam_3} \ran \lan 3, 4 \ran}\eea
The `residue' of $R_2$ at $\lan\lam_{p_{5, 7}}, \lam_{p_{1, 2}} \ran(=\lan \lam_{p_{1, 2}}, \lam_{p_{3, 4}} \ran)=0$ is
\bea  \tilde{R}_2 &=&\frac{\alpha_3^3}{\alpha_1 \alpha_2^2}A^1(-p_{1, 2}^-, 1, 2)\frac{1}{p_{1, 2}^2}A^1(-p_{5, 7}^+, 5, 6, 7) \frac{1}{p_{5, 7}^2}\frac{\lan\lam_4, \lam_{-p_{3, 4}} \ran^3 }{\lan \lam_{-p_{3, 4}, \lam_3} \ran \lan 3, 4\ran} \nonumber\\
&=&\frac{\alpha_3^3}{\alpha_1 \alpha_2^2}A^1(-p_{1, 2}^-, 1, 2)\frac{1}{p_{1, 2}^2}A^1(-p_{5, 7}^+, 5, 6, 7) \frac{1}{p_{5, 7}^2} A^1(-p_{3, 4}^-, 3, 4).\eea
The sum of the above three `residues' is
\be\sum_{i=1}^3\tilde{R}_i=\left(\frac{p_{1, 2}^2}{\alpha_1}+\frac{p_{3, 4}^2}{\alpha_2}+\frac{p_{5, 7}^2}{\alpha_3}\right)\frac{\alpha_3^3}{\alpha_1\alpha_2}A^1(-p_{1, 2}^-, 1, 2)\frac{1}{p_{1, 2}^2}A^1(-p_{3, 4}^-, 3, 4)\frac{1}{p_{3, 4}^2}A^1(-p_{5, 7}^+, 5, 6, 7) \frac{1}{p_{5, 7}^2}, \ee
which vanishes due to eq.~\eqref{eq:identity} proved in Appendix~\ref{appendix:identity}.

\section{Degenerate case\label{appendix:degenerate}}

In this appendix, we briefly display the cancelation of spurious poles for the degenerate case. We start with the pole at $\lan \lam_{n_1}, \lam_{p_{n_1+1, n_3-1}} \ran=0$ and assume that $h_{n_1}=+1$. (The case with $h_{n_1}=-1$ can be treated similarly for generic external momenta.)
This means that $n_2=n_1+1$.
Momentum conservation
\be p_{n_1}+P_2+P_3=0\,, \ee
leads to \be \lam_{n_1} \phi_{n_1}+\lambda_{P_2}+\lambda_{P_3}=0\,. \ee
From this, we get \be\lan \lambda_{n_1}, \lambda_{P_2}\ran=\lan \lam_{P_3}, \lam_{n_1} \ran
=\frac1{\phi_{n_1}}\lan \lam_{P_2}, \lam_{P_3} \ran \,. \ee
When $\lan \lam_{n_1}, \lam_{P_2}\ran=0$, we have that there should exist $v_0$ such that
\be \lam_{n_1}=\alpha_1 v_0, \lam_{P_2}=\alpha_2 v_0, \lam_{P_3}=\alpha_3 v_0\,,  \ee
with the following constraint
\be\phi_{n_1}\alpha_1+\alpha_2+\alpha_3=0\,, \ee
satisfied.

Let us consider the above pole with $h_{P_2}=+1$.
As in the non-degenerate case, we get
\be  {\text{Res}}_{\lan \lam_{n_1}, \lam_{P_2}\ran=0}V_{m_3+2}(n_1, P_2^+, q_{1(3)}^{h_{1(3)}},
\cdots, q_{m_3(3)}^{h_{m_3(3)}})=\frac{\alpha_3^2}{\alpha_1\alpha_2} V_{m_3+1} (-P_3^+, q_{1(3)}^{h_{1(3)}},
\cdots, q_{m_3(3)}^{h_{m_3(3)}})\,,\ee
which gives
\be  {\text{Res}}_{\lan\lam_{n_1}, \lam_{P_2} \ran=0} A^1(P_2^+, n_3, \cdots, n_1-1, n_1)=\frac{\alpha_3^2}{\alpha_1\alpha_2}
A^1(-P_3^+, n_3, \cdots, n_1-1)\,.\ee
So the `residue' of \be  R_3=A^1(-P_2^-, n_2,\cdots,  n_3-1)\frac{1}{P_2^2} A^1(P_2^+, n_3, \cdots, n_1-1, n_1)\,,\ee
at the pole $\lan\lam_{n_1}, \lam_{P_2} \ran=0$ is
\be \tilde{R}_3=\frac{\alpha_3^2}{\alpha_1\alpha_2} A^1(-P_2^-, n_2,\cdots,  n_3-1)\frac{1}{P_2^2}A^1(-P_3^+, n_3, \cdots, n_1-1)\,. \ee

Similarly, the `residue` of \be R_2=A^1(-P_3^+, n_3, \cdots, n_1-1)\frac1{P_3^2} A^1(P_3^-, n_1, \cdots, n_3-1)\,, \ee
at the pole $\lan \lam_{P_3}, \lam_{n_1}\ran=0$ is
\be \tilde{R}_2=\frac{\alpha_3^3}{\alpha_1\alpha_2^2} A^1(-P_2^-, n_2,\cdots,  n_3-1) A^1(-P_3^+, n_3, \cdots, n_1-1)\frac{1}{P_3^2}\,. \ee

So we have
\bea \tilde{R}_2+\tilde{R}_3&=&\frac{\alpha_3^3}{\alpha_1\alpha_2}
\left(\frac{P_2^2}{\alpha_2}+\frac{P_3^2}{\alpha_3}\right)\nonumber\\
&\times&
A^1(-P_2^-, n_2,\cdots,  n_3-1) \frac1{P_2^2} A^1(-P_3^+, n_3, \cdots, n_1-1)\frac{1}{P_3^2} . \eea
From the proof of the identity eq.~\eqref{eq:identity} in Appendix~\ref{appendix:identity}, we get
\bea \frac{P_2^2}{\alpha_2}+\frac{P_3^2}{\alpha_3}
&=&\frac1\Delta \left(\tilde{\eta}^2 \sum_{1\leq i \leq n, i\neq n_1}\tilde{\lambda}_{i1}\lan \lambda_i, v_0 \ran-\tilde{\eta}^1 \sum_{1\leq i \leq n, i\neq n_1} \tilde{\lambda}_{i2}
\lan \lambda_i, v_0 \ran \right)\nonumber\\
&=&-\frac1\Delta \left(\tilde{\eta}^2 \tilde{\lambda}_{n_1\, 1}\lan \lambda_{n_1}, v_0 \ran-\tilde{\eta}^1\tilde{\lambda}_{n_1\, 2}
\lan \lambda_{n_1}, v_0 \ran \right)=0,
 \eea
 where we have used that $\lambda_{n_1}$ is proportional to $v_0$ at the pole. So finally we get $\tilde{R}_2+\tilde{R}_3=0$.
 Other cases can be treated in the same way.

\section{Proof of eq.~\eqref{eq:identity}\label{appendix:identity}}
From \cite{Zhu:2004kr},  we get
\be\label{eq:Zhu} P_1^2=p_{n_1, n_2-1}^2=\sum_{i=n_1}^{n_2-1} \frac{[i, k]}{\phi_i \phi_k} \lan \lambda_{p_i}, \lambda_{p_{n_1, n_2-1}}\ran\,,  \ee
with $k$ can be any integer satisfying $1\leq k \leq n$.

Recall that
\be \phi_i=\tilde{\lambda}_{i\dot{\alpha}} \tilde{\eta}^{\dot{\alpha}}=\tilde{\lambda}_{i1}\tilde{\eta}^1+\tilde{\lambda}_{i2}\tilde{\eta}^2\,,\ee
let us further define
\be \psi_i\equiv \tilde{\lambda}_{i1}\tilde{\eta}^2-\tilde{\lambda}_{i2}\tilde{\eta}^1\,, \, 1\leq i\leq n. \ee

From the above two equations, we can express $\tilde{\lambda}_{i\dot{\alpha}}$ in terms of $\phi_i, \psi_i, \tilde{\eta}^{\dot{\alpha}}$ as
\bea
\tilde{\lambda}_{i1}&=&\frac{\tilde{\eta}^1 \phi_i+\tilde{\eta}^2 \psi_i}{(\tilde{\eta}^1)^2+(\tilde{\eta}^2)^2}\,,\\
\tilde{\lambda}_{i2}&=&\frac{\tilde{\eta}^2 \phi_i-\tilde{\eta}^1 \psi_i}{(\tilde{\eta}^1)^2+(\tilde{\eta}^2)^2}\,.
\eea
Then we  get
\bea [i, k]=\tilde{\lambda}_{i1}\tilde{\lambda}_{k2}-\tilde{\lambda}_{i2}\tilde{\lambda}_{k1}=
\frac{\psi_i\phi_k-\psi_k\phi_i}{(\tilde{\eta}^1)^2+(\tilde{\eta}^2)^2}\,,
 \eea
and so \be\frac{[i, k]}{\phi_i\phi_k}=\frac{1}{(\tilde{\eta}^1)^2+(\tilde{\eta}^2)^2}\left(\frac{\psi_i}{\phi_i}-\frac{\psi_k}{\phi_k}\right)\,. \ee
Now we define
\be \Delta\equiv (\tilde{\eta}^1)^2+(\tilde{\eta}^2)^2,\, \varphi_i=\frac{\psi_i}{\phi_i},\, 1\leq i\leq n,\ee
we have \be \frac{[i, k]}{\phi_i\phi_k}=\frac{\varphi_i-\varphi_k}{\Delta}\,. \ee
From this result and \eqref{eq:Zhu}, we obtain
\be P_1^2=\frac1{\Delta} \sum_{i=n_1}^{n_2-1} (\varphi_i-\varphi_k) \lan \lambda_{p_i}, \lambda_{p_{n_1, n_2-1}} \ran\,. \ee
Obviously
\bea \sum_{i=n_1}^{n_2-1}\varphi_k \lan \lambda_{p_i}, \lambda_{p_{n_1, n_2-1}}\ran=\varphi_k \lan \sum_{i=n_1}^{n_2-1}\lambda_{p_i}, \lambda_{p_{n_1, n_2-1}}\ran =0 \,.
 \eea
So \be P_1^2=\frac1{\Delta}\sum_{i=n_1}^{n_2-1} \varphi_i \lan \lambda_{p_i}, v_1\ran \ee
Then \bea \frac{P_1^2}{\alpha_1}&=&\frac1{\Delta} \sum_{i=n_1}^{n_2-1}\varphi_i \lan \lambda_{p_i}, v_0 \ran\nonumber\\
&=& \frac1\Delta \sum_{n_1}^{n_2-1} \varphi_i \phi_i\lan\lambda_i, v_0 \ran\nonumber\\
&=&\frac1\Delta \sum_{n_1}^{n_2-1}\psi_i \lan \lambda_i, v_0 \ran\,,
  \eea
  where we have used $v_1=\alpha_1 v_0$ and $\varphi_i=\frac{\psi_i}{\phi_i}$.

Similarly, we have
\bea \frac{P_2^2}{\alpha_2}&=&\frac1\Delta  \sum_{i=n_2}^{n_3-1} \psi_i \lan\lambda_i, v_0\ran, \\
\frac{P_3^2}{\alpha_2}&=&\frac1\Delta  \sum_{i=n_3}^{n_1-1} \psi_i \lan\lambda_i, v_0\ran,
\eea
From these results, we get
\bea  \frac{P_1^2}{\alpha_1}+\frac{P_2^2}{\alpha_2}+\frac{P_3^2}{\alpha_3} &=&\frac{1}{\Delta}\sum_{i=1}^n \psi_i \lan \lambda_i, v_0 \ran \nonumber\\
&=&\frac1\Delta \sum_{i=1}^n\left(\tilde{\lambda}_{i1}\tilde{\eta}^2-\tilde{\lambda}_{i2}\tilde{\eta}^1\right)\lan\lambda_i, v_0  \ran
\nonumber\\
&=&\frac1\Delta \left(\tilde{\eta}^2 \sum_{i=1}^n \tilde{\lambda}_{i1}\lan \lambda_i, v_0 \ran-\tilde{\eta}^1 \sum_{i=1}^n \tilde{\lambda}_{i2}
\lan \lambda_i, v_0 \ran \right)\nonumber\\
&=&0\,,
 \eea
where momentum conservation $\sum_{i=1}^n \lambda_i \tilde{\lambda}_i=0$ has been used. This finishes the proof.

\section{Mass dimensions of $A_{\text{CSW}}$ and $A_{\text{Feynman}}$ \label{appendix:dimensions}}
We would like to show explicitly that both $A_{\text{CSW}}$ and $A_{\text{Feynman}}$ has mass dimension $4-n$.
We first  show that the mass dimension of $A_{\text{CSW}}$ is $4-n$.
For an $n$-gluon tree amplitude, assume $n_{\pm}$ being the number of external gluons with helicity $\pm 1$, $n_p$ being the number of propagators, and $n_i$ being the number of MHV vertices with exactly $i$ lines. From \cite{Cachazo:2004kj, Zhu:2004kr}, we get
\bea \label{nminus} n_-&=&\sum_i n_i+1,\\
\label{nplus}n_+&=&\sum_i n_i (i-3)+1. \eea
And it is easy to see that \be \label{np}n_p=\sum_i n_i-1=n_--2. \ee
Notice that spinors $\lambda, \tilde{\lambda}$ of external gluon have dimension  $[M]^{1/2}$.
Then the $m-$gluon MHV amplitudes has mass dimension $4-m$.
Since $A_{\text{CSW}}$ is invariant under $\tilde{\eta}^{\dot{\alpha}}\to t\tilde{\eta}^{\dot{\alpha}}$, we can assign $\tilde{\eta}^{\dot{\alpha}}$ arbitrary dimension. It is convenient to assign its dimension to be $[M]^{-1/2}$, such that $\lambda$ for the internal line has the same dimension as $\lambda$ for the external gluon. Then the MHV vertex with  $i$ lines has mass dimension $4-i$, the same as the one of $i-$gluon MHV amplitudes. Also notice that every propagator has mass dimension $-2$. So the mass dimension of the contribution of an MHV diagram to the amplitude is
\be\left(\sum_i(4-i)n_i\right)-2 n_p \ee
which can be shown to be $4-n_+-n_-=4-n$ by using \eqref{nminus}-\eqref{np}.

The fact that $A_{\text{Feynman}}$ has mass dimension $4-n$ can be proved similarly. Denote the number of vertices with $i$ lines by $\tilde{n}_i, i=3, 4$ and the number of propagators by $\tilde{n}_p$.  We know that
\bea \tilde{n}_p&=&\tilde{n}_3+\tilde{n}_4-1,\\
n&=&3\tilde{n}_3+4\tilde{n}_4-2\tilde{n}_p.\eea
From these two equations, we get
\bea \tilde{n}_3&=&n-2\tilde{n}_4-2,\\
\tilde{n}_p&=&n-\tilde{n}_4-3. \eea
Then the mass dimension of a Feynman diagram is\footnote{Notice that the polarization vectors in eq.~\eqref{eq:polarizations} are dimensionless.}
\be \tilde{n}_3-2\tilde{n}_p=4-n\,. \ee



	\end{document}